\documentstyle[amssymb,aps,prb,twocolumn,psfig]{revtex}

\draft

\begin{document}

\twocolumn[\hsize\textwidth\columnwidth\hsize\csname@twocolumnfalse\endcsname

\title{Magnetic polarons in $Ca_{1-x}Y_{x}MnO_{3}$}
\author{H. Aliaga, M. T. Causa, M. Tovar and B. Alascio}
\address{Comisi\'{o}n Nacional de Energ{\'{\i }}a At\'{o}mica,\\
Centro At\'{o}mico Bariloche and Instituto Balseiro, 8400 S.C. de Bariloche,%
\\
Argentina}

\maketitle

\begin{abstract}
Experimental evidence show that in the magnetoresistive manganite $Ca_{1-x}
Y_{x}MnO_{3}$, ferromagnetic (FM) polarons arises in an antiferromagnetic
(AF) background, as a result of the doping with Yttrium. This hypothesis is
supported in this work by classical Monte Carlo (MC) calculations performed
on a model where FM Double Exchange (DE) and AF Superexhange (SE) compite.
\end{abstract}

]

In this work we try to explain experimental results\cite{aliaga} on
magnetoresistive $Ca_{1-x}Y_{x}MnO_{3}$ doped manganites, that point towards
the existence of magnetic polarons. The pure sample ($x$=0) presents G-type AF
ordering with a transition temperature $T_{N}$=123K, and insulator
characteristics. In this sample, all the Mn ions have valence 4+, where the
localized $t_{2g}$ spins have a spin $S$=3/2. As Yttrium ions are added,
itinerants $e_{g}$ electrons with spin $s$=1/2 are introduced and there is a
mixed valence state of $Mn^{4+}$ and $Mn^{3+}$ in the proportion $(1-x)$ and 
$x$, respectively. The magnetic moment raises\cite{aliaga} and the
resistivity drops sharply as $x$ is increased up to $x\leq $0.15.
A thermally activated polaronic behaviour for the conductivity was fitted between
room temperature and 100K. The experimental results were interpreted in terms
of FM polarons, produced by the alignment of neighboring $t_{2g}$ spins by
the $e_{g}$ electrons and immersed in the AF background. In
this work we compare the experimental magnetic measurements with the
properties predicted by MC calculations on a model where FM-DE and AF-SE
interactions compete.

\begin{figure}[t]
\centerline{\psfig{figure=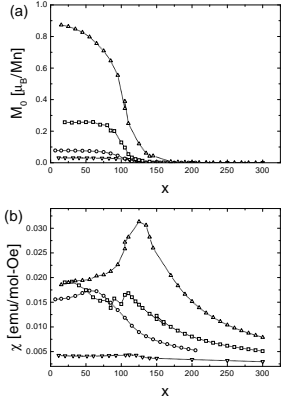,height=6cm,angle=0}}
\medskip
\caption{(a) $M_{0}$ vs. $T$, and (b) $\chi$ vs. $T$, for the samples $x$%
=0 (down triangles), 0.03 (circles), 0.05 (squares), and 0.1 (up triangles).}
\end{figure}

Ceramic polycrystalline samples of $Ca_{1-x}Y_{x}MnO_{3}$ were prepared by
solid state reaction methods\cite{cac}. Room temperature x-rays
diffractograms show that all the samples crystallize in an orthorhombic $Pnma
$ cell. The dc-magnetization, $M$, was measured with a SQUID magnetometer
between 5K and 300K for $H\leq $50kOe.

We measured $M$ vs. $H$ at different temperatures for the samples $x$=0,
0.03, 0.05 and 0.1 and we found a linear dependence at high fields: $M$=$M_{0}$+%
$\chi $ $H$. In Fig. 1(a) we show $M_{0}$ vs. $T$. The behavior resembles
that of a ferromagnet with a characteristic ordering temperature $T_{mo}$,
almost constant for all the samples. For $x$=0, a small $M_{0}\sim$0.03$\mu
_{B}$/f.u. reflects a weak ferromagnetic Dzialoshynskii-Moriya
interaction\cite{aliaga}. In Fig. 2(a) we plot $M_{0}$ vs. $x$ at $T$=5K. The curve has
an $S$ shape starting with a slope of 1$\mu _{B}$ per $Mn^{3+}$ ion and 
reaching a slope $M_{0}$/$x$$\sim$7$\mu _{B}$ at $x$=0.07. 
Notice that full FM alignment (3$\mu _{B}$/Mn site) is not reached.

\begin{figure}[t]
\centerline{\psfig{figure=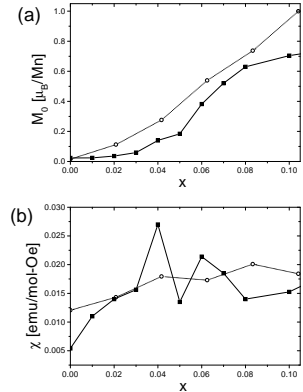,height=6cm,angle=0}}
\medskip
\caption{(a) $M_{0}$ and (b) $\chi$, as function of doping $x$, measured
at 5K (squares), and calculated (circles).}
\end{figure}

In Fig. 1(b), we show $\chi (T)$ vs. $T.$ The general behavior above $T_{mo}$
has been studied in Ref. 1, for $x$=0, 0.05, 0.07 and 0.1. The most peculiar
feature observed is a deviation from the Curie-Weiss law, presenting a
negative curvature in $\chi ^{-1}(T).$ This temperature dependence was
described in terms of two contributions:

\begin{eqnarray}
\chi (T)=(1-x)\chi _{AF}(T)+x\text{ }\chi _{FM}(T), 
\end{eqnarray}

where $\chi _{AF}(T)$ is the susceptibility of the AF background and $\chi
_{FM}(T)$ reflects the strong FM-DE correlations that results in polaron
formation in the PM phase. For $x$=0, a maximum in $\chi(T)$ is present at $%
T_{N}$. For $x$=0.1 the behavior is qualitatively similar: a maximum is 
observed around $T_{mo}$, although with a much larger value. Below $T_{mo}$, 
$\chi(T)$ decreases rapidly and reaches $\sim0.6\chi(T_{mo})$ at $T$=5K. 
For the intermediate concentration, $x$=0.05, a peak at $T_{mo}$
is only a relative maximum: $\chi (T)$ increases again below $T$$\sim$70K
and $\chi _{AF}(5K)\sim1.2\chi (T_{mo})$. For the low doped $x$=0.03 sample
no peak is observed at $T_{mo}$ and $\chi (T)$ continues to increase below $%
T_{mo}$ up to $T$$\sim$60K, where a broad maximum is present. The large
values of $\chi (T_{mo})$ take into account the progressive polaron
formation, indicated by the proportionality $\chi (T_{mo})$ $\varpropto $ $x$,
given by Eq.(1), where $\chi_{AF}(T_{mo})\ll\chi_{FM}(T_{mo})$. 
In Fig. 2(b), we show $\chi(5K)$ vs. $x$

The simplest picture for the behavior of $M$($H$,$T$) below $T_{mo}$, is to
associate $\chi (T)$ with the AF background and assume that $%
M_{0}$($T$) corresponds to the FM order of polarons. However, the measured 
$\chi (T)$ is
in all cases much larger than the $CaMnO_{3}$ susceptibility. Thus, this
simple model seems not appropriate to describe the experiments in the ordered
region. In order to obtain a better description of the magnetization
behavior, we have performed a calculation of $M$ vs. $H$ using the numerical
classical MC technique.

We consider the following hamiltonian: 
\begin{eqnarray}
{\cal H} = -\sum_{\langle ij\rangle \sigma }t_{ij}\left( c_{i\sigma }^{\dag
}c_{j\sigma }+h.c.\right) -J_{H}\sum_{i}{\bf s}_{i}{\bf .S}_{i} 
\nonumber \\ +g\mu_{B}H\sum_{i}S_{z,i}+K\sum_{\langle ij\rangle }{\bf S}_{i}{\bf .S}_{j} , 
\end{eqnarray}
Here $c_{i\sigma }^{\dag }$ is the operator creating an itinerant electron
of spin $\sigma $ at site i, and ${\bf s}_{i}=\sum_{\alpha \beta }c_{i\alpha
}^{\dag }\sigma _{\alpha \beta }c_{i\beta }$ gives the spin projection of the electron.
Due to the strong Hund coupling ($J_{H}\rightarrow\infty$) 
between itinerants $e_{g}$ electrons and the localized $t_{2g}$
spins, only parallel spin projections were taken into account. In this way,
the hopping of itinerant electrons depends on the orientation of the
localized $t_{2g}$ spins, according to the DE expression in one dimension: $%
t_{ij}$=$t$ cos($\theta _{ij}/2$), where $\theta _{ij}$ is the relative
angle between localized spins {\bf $S_{i}$} and {\bf $S_{j}$}, $t$ is the
hopping parameter. The third term, represents Zeeman coupling between
magnetic field $H$ and $S_{z,i}$, the z component of localized spin {\bf $%
S_{i}$}. The AF-SE interaction between localized neighboring spins {\bf $%
S_{i}$} and {\bf $S_{j}$} is represented by the last term in Eq.(2).
Numerical calculations were performed in one dimension, using the MC
algorithm for a 48 sites chain with open boundary
conditions. The localized spins were classical with modulus 1, and later
renormalized to (3+$x$)$\mu_{B}$ in order to compare with experiments. Local changes in $t_{2g}$
spins were made in conjunction with exact diagonalization of the itinerant
electron system. The resulting electronic energy levels were then filled by
the available number of electrons in the canonical ensemble. We take the 
hopping parameter t=0.1eV as the reference energy, and the temperature $T$=0.005t$\sim$6K.
Mermin-Wagner's theorem precludes magnetic ordering
at finite temperatures in 1D systems. However, we estimated a value for $K$
from the ordering temperature $T_{mo}$ that would arise from a classical
Heisenberg model for a system with $z$=2 neighbors. For $T_{mo}$$\sim $100K,
we derive $KS^{2}/t$=0.05.

\begin{figure}[t]
\centerline{\psfig{figure=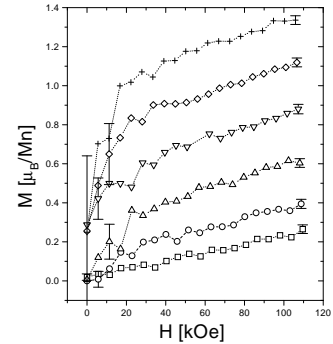,height=6cm,angle=0}}
\medskip
\caption{Magnetization $M$ vs. $H$, for $x$=0(squares), 1/48(circles),..
,5/48(plus signs), obtained by MC calculations. Characteristics error bars 
are displayed for few points.}
\end{figure}

In Fig. 3, we show $M$ vs. $H$ dependence for the 1D chain, when 0,1,..,5
electrons are added (equivalent to dopings $x$=0, 0.02,..,0.1). For the
model considered, there is a competence between AF-SE and FM-DE
interactions. Preliminary calculations\cite{unp} show that a polaronic phase
is obtained for $K/t\geqslant $0.6 $x$. These polarons are immersed in the
otherwise AF background. As we apply a magnetic field, the FM\ polarons tend
to align in the direction of $H$. We calculated $M(T,x,H)$ averaging over 5000 MC
realizations and sites. $M_{0}$ was obtained extrapolating a linear
dependence from the highest fields (40kOE$\le H\le$120kOe). When all polarons are aligned, the ratio 
$M_{0}$/$x$ gives a measurement of the polaron size. The high field
differential slope of $M$ vs. $H$ calculated in this case would then arise
from two main processes: enlargement of the polarons and canting of the AF
background.

The slope at high fields is of the same order in all the samples. The linear
response in the case $x$=0, is due only to the canting of the AF background, 
$\chi (x=0)\varpropto K^{-1}$. For $x>$0, in the range $H\leq $75kOe, we
find a rapid growth of $M$ vs. $H$, although strong fluctuations are present 
due to finite-size effects. In this regime the polarons, originally randomly oriented,
start to align with $H$\cite{aliaga}. The numerical results suggests that full alignment
is not completely reached at 50kOe, which is our experimental limit.

In Fig. 2(a), we compare the $M_{0}$ vs. $x$ dependence obtained
numerically, with the experimental values at 5K. As we mentioned
above, the slope of $M_{0}$ vs. $x$ is a measurement of the magnetic moment
of the polaron created by each $e_{g}$ electron added. In the experiments,
we find for $x<$0.03, that each polaron has a magnetic moment of 1$\mu _{B}$%
, and for $x>$0.03, increases up to 7$\mu _{B}/$polaron$.$ In the present
model, we find a linear dependence in the curve $M_{0}$ vs. $x$. The size of
each polaron is fixed by the ratio $K/t$\cite{unp}. The distribution of the 
itinerant charge, shows that each polaron extends over 5 to 6 sites.
The differential susceptibility $%
\chi $ is a measurement of the the canting of the AF background and 
the enlargement of polarons with $H$. In Fig. 2(b) we show the calculated 
$\chi $ vs. $%
x$, that shows an almost constant behaviour. Since for $x$=0.1, about half 
of the sites take part in the polaron
phase, we conclude that the
response of polarons to $H$ is of the same order to that of the canting of the AF
background.  

We found a remarkable agreement between experimental and
numerical values of $M_{0}$ and $\chi $. The numerical simulations 
suggest that linear regime $M$ vs. $H$, is not fully reached at $H$=50kOe, 
thus experiments at higher fields are suggested in order to make a better 
comparison with numerical prediction, specially in the low doped regime, 
where some discrepancies are observed.

We acknowledge partial support from ANPCYT (Argentina)-PICT 3-52-1027/3-05266, and CONICET(Argentina)-PIP 4947/96. H.
A. is CONICET (Argentina) PhD-fellow.

\end{document}